\documentclass{cs19proc}

\usepackage{kantlipsum}

\editors{G.~A. Feiden}
\publisher{Zenodo}
\conference{The 19th Cambridge Workshop on Cool Stars, Stellar Systems, and the Sun}
\conferencedate{2016}

\title{On the bimodal distribution of stellar rotation in young open clusters}
\author{Philippe Gondoin}

\affiliation{European Space Agency, ESTEC - Postbus 299,
              2200 AG Noordwijk, The Netherlands}

\shorttitle{Stellar rotation in clusters}
\shortauthors{Ph. Gondoin}

\abs{Observations of young open clusters show a bimodal distribution of stellar rotation. Sun-like stars in those clusters group into two main sub-populations of fast and slow rotators. Beyond an age of about 500 Myrs, the two populations converge towards a single peak distribution of angular velocities. I argue that this evolution of stellar rotation in open clusters results from a brief episode of enhanced angular momentum loss by strong stellar wind during the early evolution of rapidly rotating Sun-like stars.}

\begin{document}

\maketitle

\section{Introduction}

In the past decades, photometric survey of open clusters have produced extensive rotation period measurements on stars with different ages. The results indicate a spin-up phase of stellar rotation during pre-main sequence contraction followed by a spin-down near the ZAMS, and during further evolution on the main sequence. 

These measurements also show (e.g. Barnes 2003; Meibom et al. 2009; 2011) a bimodal distribution of stellar rotation in young open clusters. Young Sun-like stars tend to group into two distinct populations of slow and fast rotators. These populations lie on narrow sequences in diagrams where the measured rotation periods of the members of a stellar cluster are plotted against their B $-$ V colors. One sequence consists of stars that form a diagonal band of increasing rotation period with increasing B - V color. In young clusters, another sequence of fast rotators is also observed. Few stars lie in the intervening gap between these two sequences. Beyond the age of about 500 Myrs, the two groups of fast and slow rotators converge towards a single distribution of angular velocities.

Modelling the evolution of stellar rotation has been the subject of many studies. However, very few papers (Brown 2014) have intended to account for the bimodal distribution of stellar rotation observed in young open cluster. The present work addresses the origin and evolution of this bimodal distribution. Section 2 describe a phenomenological model of stellar rotation. Section 3 present simulation results of the rotation evolution a Sun-like star as a function of its initial rotation rate. It compares the evolution of a normal distribution of stellar rotation periods with measurements of stellar rotation in open clusters of various ages. The results are discussed in Sect. 4.

\section{Model of stellar rotation evolution}

Studies of the rotation evolution of Sun-like stars often use a phenomenological model of angular momentum redistribution in stellar interiors, the so-called Double Zone model (DZM).  This model assumes that the radiative core and the convective envelope rotate rigidly at all ages. These hypotheses are justified (Spada et al.  2011) by the facts that (i) the envelope is expected to be strongly coupled by the very efficient angular momentum redistribution associated with turbulent convection, and (ii) a large-scale magnetic field, even as weak as 1 G, can maintain a condition of rigid rotation in most of the core.  

At any given time, t, the angular momenta J$_{\rm c}$ of the core and J$_{\rm e}$ of the envelope can thus be expressed as $J_{\rm c}(t) = I_{\rm c}(t) \Omega_{\rm c}(t)$ and $J_{\rm e}(t)= I_{\rm e}(t) \Omega_{\rm e}(t)$.  I$_{\rm c}$, I$_{\rm e}$, $\Omega_{\rm c}$, $\Omega_{\rm c}$ are the moments of inertia and the angular velocities of the core and of the envelope, respectively.  Assuming that the angular momentum redistribution between the two regions occurs on a timescale $\tau_{\rm c}$,  the evolution of their angular momenta is governed by two coupled differential equations (e.g Oglethorpe \& Garaud 2013):
\begin{equation}
	{dJ_{\rm c} \over dt} =  -{\Delta J \over \tau_{\rm c}} + \bigg({2 \over 3} R_{\rm c}^{2} {dM_{\rm c} \over dt}\bigg)\Omega_{\rm e} 
\end{equation}
\begin{equation}	
	{dJ_{\rm e} \over dt} = {\Delta J \over \tau_{\rm c}} - \bigg({2 \over 3} R_{\rm c}^{2} {dM_{\rm c} \over dt}\bigg)\Omega_{\rm e}  -  \tau_{\rm w} 
\end{equation}
where 
\begin{equation}
	\Delta J = {I_{\rm c}I_{\rm e} \over I_{\rm c} + I_{\rm e}}(\Omega_{\rm c} - \Omega_{\rm e}).
\end{equation}
$\tau_{\rm w}$ is the torque exerted by the wind on the convective envelope.

During the pre-main sequence evolutionary phase, the growth of the core at the expenses of the envelope produces an angular momentum transfer, accounted for by the second term in the right-hand side of both equations (1) and (2). $R_{\rm c}$ and $M_{\rm c}$ are the radius and mass of the radiative core. These stellar quantities and their time derivatives are taken from the evolutionary track of a solar mass star established by Siess et al. (2000) with a metallicity $Z$ = 0.02 and a moderate overshoot parameter $d$ = 0.20 $H_{\rm p}$  where $H_{\rm p}$ is the pressure scale height.  

Weber \& Davis (1967) demonstrated that the loss rate of angular momentum by the star is expressed in a steady state as the product of the mass loss rate and the specific angular momentum carried by the outflow, i.e.:

\begin{equation}
\tau_{\rm w} =  \Omega_{\rm e} \dot{M_{\rm w}}  R_{\rm A}^{2}
\end{equation}

where $\dot{M_{\rm w}}$ is the mass loss rate and $R_{\rm A}$ the Alfven radius. 

According to the simple model described above, the angular momentum evolution of a Sun-like star depends on the time evolution of its internal structure and on three parameters describing the angular momentum loss and its redistribution within the stellar interior. These parameters are the mass loss rate $\dot{M_{\rm w}}$, the mean Alfven radius $R_{\rm A}$, and the core-envelope coupling timescale  $\tau_{\rm c}$. 

The core-envelope coupling timescale $\tau_{\rm c}$ was initially found to be constant throughout the evolution of a 1 M$_{\rm \odot}$ star with a magnitude ranging from 10 Myrs  to a few times 10 Myrs. In order to account for new observational results, recent studies suggested that the coupling timescale may be different for fast and slow rotators. Based on the argument that the derived dependency is weak, the present study assumes that the core-envelope coupling timescale of a Sun-like star is constant over time and independent of its initial rotation rate.

Observations (Wood et al. 2005, 2014) indicate that the mass loss rate of Sun-like stars increases as a power law of their X-ray surface flux. The interpretation of theses observations with the assistance of hydrodynamic models also suggest that, beyond a certain flux,  the mass loss rate of young stars drops to weaker values. Since the X-ray luminosity and therefore the X-ray surface flux of cool stars is a function of their Rossby number, I formulated this scaling law as follows:
\begin{equation}
\dot{M_{\rm w}} \approx \left\{ \begin{array}{rl}
\dot{M_{\rm \odot}} \times (F_{\rm X}/ F_{\rm X, \odot})^{\alpha} &\mbox{if $Ro >$ Ro$_{\rm w}$} \\ 
\dot{M_{\rm sat}}  &\mbox{if $Ro \le$ Ro$_{\rm w}$} \\
       \end{array} \right.
\end{equation}
where  $F_{\rm X}$ and $F_{\rm X\odot }$ are the stellar and solar X-ray surface flux, respectively. Ro$_{\rm w}$ is the Rossby number at a so-called wind dividing line below which the mass loss rate saturates at a constant value $\dot{M_{\rm sat}}$. 

The X-ray to bolometric luminosity ratio of late-type stars has been parameterised as a function of the Rossby number Ro (Pizzolato et al. 2003):
\begin{equation}
{L_{\rm X} \over L_{\rm bol}} \approx \left\{ \begin{array}{rl}
 R_{\rm X, sat} &\mbox{if $Ro \le Ro_{\rm crit}$} \\
 (L_{\rm X, \odot} / L_{\rm \odot})(Ro / Ro_{\rm \odot})^{\beta}  & \mbox{if $Ro >$ Ro$_{\rm crit}$.}
       \end{array} \right.
\end{equation}
with $\log(L_{\rm X,\odot}/L_{\rm \odot})$ = -6.24 (Judge et al. 2003). In this equation, $L_{\rm bol}$ is the stellar bolometric luminosity and $Ro_{\rm crit}$ is the Rossby number below which the saturation of X-ray emission occurs.  R$_{\rm X, sat}$ = 0.74 $\times$ 10$^{-3}$ is the saturation level of the X-ray to bolometric luminosity ratio. Wright et al. (2011) found a power index $\beta$ = -2.7.  I combined Eq. (5) and (6) to parameterise the mass loss rate of Sun-like stars as a function of their Rossby number.

One observational constraint on the rotation evolution of Sun-like stars comes from the well known result of Skumanich (1972) that stars older than about 1 Gyr spin down with the inverse of the square root of time. This rotation evolution imposes $R_{\rm A} \approx$  R$_{\rm A, \odot }  \times \left\{\Omega_{\rm e} / \Omega_{\rm \odot}\right\}^{1+{\alpha \beta \over 2}}$ for slow rotators.  Assuming that the Alfven radius of rapid rotators saturates at Rossby number smaller than Ro$_{\rm A}$, $R_{\rm A}$ can thus be expressed as follows:
\begin{equation}
R_{\rm A} = \left\{ \begin{array}{rl}
R_{\rm A, \odot } \left\{\Omega_{\rm e} / \Omega_{\rm \odot}\right\}^{1+{\alpha \beta \over 2}} & \mbox{if $Ro >$ Ro$_{\rm A}$}\\ 
R_{\rm A, \odot } \left\{P_{\rm \odot} /( \tau_{\rm \odot} . Ro_{\rm A}) \right\}^{1+{\alpha \beta \over 2}} &\mbox{if $Ro \le Ro_{\rm A}$} \\     
	 \end{array} \right.
\end{equation} 
where  $\tau_{\rm \odot}$ and P$_{\rm \odot}$ are the convective turnover time and the rotation period of the Sun, respectively.

\section{Simulations}

\subsection{Rotation evolution of a Sun-like star}

In order to simulate the rotation evolution of one solar mass stars, I first used the empirical parameterisations of the mass loss rate vs X-ray surface flux proposed by Wood et al. (2014) and that of the X-ray luminosity vs Rossby number formulated by Wright et al. (2011).  When imposing those relationships, the rotation evolution model has only three free parameters, namely the initial rotation period after circumstellar disk dispersion, the core-envelope coupling timescale, and the Rossby number Ro$_{\rm A}$ below which the Alfven radius saturates at its lowest value. For various distribution of initial rotation periods,  I could not find a set of free parameters that lead to simulation results similar to the measured distribution of rotation periods in open clusters. In particular, the simulations could not reproduce the bi-modal distribution of stellar rotation observed in intermediate age open clusters.  

I thus let the mass loss rate parameter at saturation  $\dot{M_{\rm sat}}$ vary since this parameter is poorly constrained by observation. I also imposed that the saturation of the wind occurs at Ro$_{\rm w}$ = 0.13 and that the saturation of the Alfven radius occurs at Ro$_{\rm A}$ = 0.4.  These constraints are suggested by the observations that the transition from the fast to the slow rotation sequence in the M34 open cluster and in the Pleiades occurs on stars with 0.13 $<$ Ro $<$ 0.4 (Gondoin 2014, 2015). 

\begin{figure}[!t]
	\centering
	\includegraphics[width=0.75\linewidth, angle=270]{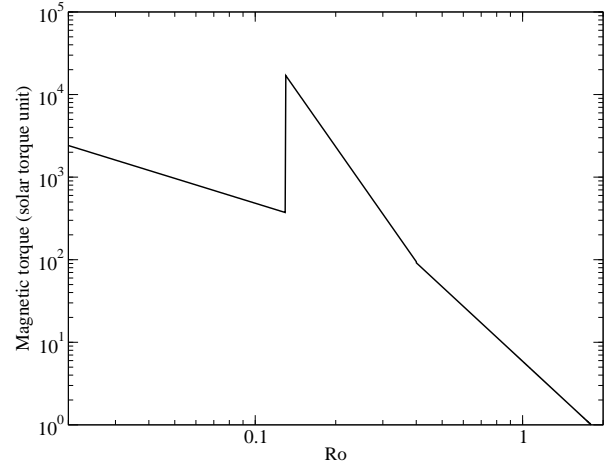} 
	\caption{Assumed braking torque  of a Sun-like star as a function of its Rossby number.}
	\label{fig:fig_sim_mag_torque}
\end{figure}

Figure 1 shows the braking torque of a 1 M$_{\rm \odot }$ star as a function of its Rossby number resulting from these hypothesis and assuming  $\dot{M_{\rm sat}}$ = 300 $\dot{M_{\rm \odot}}$. Using an average solar mass loss rate $\dot{M_{\rm \odot}}$ = 2.5 $\times$10$^{-14}$  M$_{\rm \odot }$/yr,  an Alfven radius R$_{\rm A, \odot }$ = 7.0 R$_{\rm \odot }$ leads to rotation periods in the range 22-28 days at an age of 4.5 Gyrs (i.e. similar to the Sun) for initial periods ranging from 0.6 to 30 days at 5 Myrs. Table 1 lists the model parameters that I used to simulate the angular momentum evolution of a solar mass star. 

Figure 2 plots the results for initial rotation periods ranging from 0.6 (upper curve) to 30 days (lower curve) at an age of 5 Myrs. It shows a spin-up phase during pre-main sequence contraction  followed by a spin-down near the ZAMS and during further evolution on the main sequence. Measurements of stellar angular velocities in several open clusters (see Table 2) are also plotted in Fig. 2. The simulated angular velocity curves fit reasonably well the dispersion of rotation periods among solar-type stars in young open clusters. The narrowing of their distribution between the age of M37 ($\sim$ 550 Myrs) and that of the Sun is also reproduced. 

\begin{table}[!t]
	\centering
	\caption{Model parameters used to simulate the angular momentum evolution of solar mass stars.}
	\label{tab:table_model_param}
	\begin{tabular}{c | c | c}
	\noalign{\smallskip}\hline\hline\noalign{\smallskip}
	Process & Parameter & Value \\
	\noalign{\smallskip}\hline\noalign{\smallskip}
	Coupling timescale &  $\tau_{\rm c}$ & 30 Myrs \\
	\noalign{\smallskip}\hline\noalign{\smallskip}  
	 & $\dot{M_{\rm sat}}$ & 300 $\dot{M_{\rm \odot }}$ \\
	Mass loss rate & Ro$_{\rm w}$ & 0.13 \\
 	& $\alpha$ & 1.34 \\ 
	\noalign{\smallskip}\hline\noalign{\smallskip}
	 & R$_{\rm X, sat}$ & 0.74 $\times$ 10$^{-3}$  \\ 
	X-ray luminosity  & Ro$_{\rm crit}$ & 0.13  \\
 	& $\beta$ & -2.7  \\  
	\noalign{\smallskip}\hline\noalign{\smallskip}
	& R$_{\rm A, sat }$ & 2.1  R$_{\odot }$ \\
	Alfven radius & Ro$_{\rm A}$ & 0.4  \\
 	& R$_{\rm A, \odot }$ & 7 R$_{\odot }$ \\	
	\noalign{\smallskip}\hline
	\end{tabular}
\end{table}

\begin{figure}[!t]
	\centering
	\includegraphics[width=0.75\linewidth, angle=270]{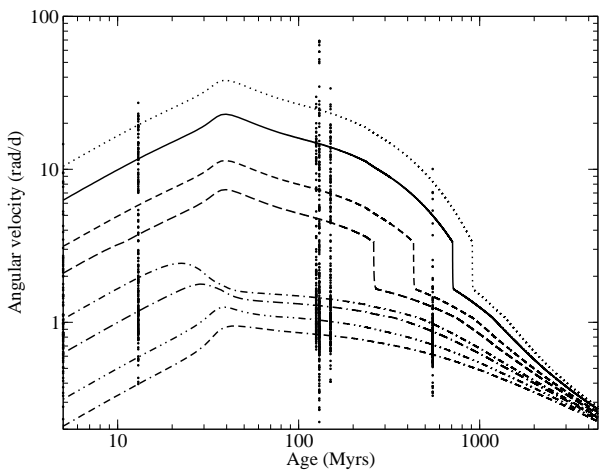} 
	\caption{Simulated angular velocity evolution of a solar mass star with initial rotation periods (at 5 Myrs age) of 0.6 (upper curve), 1, 2, 3, 6, 10, 20 and 30 days (lower curve). The simulated angular velocity curves are compared with rotation measurements in h Per (13 Myrs; Moraux et al. 2013),  the Pleiades (125 Myrs; Hartman et al. 2010), M50 (130 Myrs; Irwin et al. 2009), M35 (133 Myrs; Meibom et al. 2009), and  M37 (550 Myrs; Hartman et al. 2009) .}
	\label{fig:fig_sim_rot_evol}
\end{figure}

\subsection{Evolution of stellar rotation in open clusters}

The time series photometric survey of  NGC 2362 by Irwin et al. (2008) provided a measurement of an early distribution of rotation periods in a $\sim$  3-4 Myrs old open cluster (Mayne et al. 2007).  I extracted a sample of 91 stars with masses included between 0.7 and 1.1 M$_{\odot}$ stars from the list of Irwin et al. (2008). This sample shows a broad distribution of rotation periods between 0.4 and 25 days with a maximum around 7 days (see upper left plot in Fig.5). The histogram shows no evidence of a bimodal behavior with distinct populations of slow and fast rotators. 

Only a few stars in the parent sample show mid-infrared excesses indicative of the presence of circumstellar disks (Irwin et al. 2008). The disk accretion process has ceased for most of the stars and will not affect the subsequent evolution of their angular momentum. The measured distribution of stellar angular velocities among the 91 sample stars thus provides an example of a distribution of stellar rotation in an open cluster just after dispersion of the circumstellar disks.  

This sample however has a relatively small size compared with those derived from some surveys of older clusters (see Table 2).  I thus emulated a larger distribution of stellar angular velocities in NGC 2362 by generating a normal distribution of 1000 rotation periods with a maximum around 7 days. The distribution was truncated below 0.3 day to mimic the distribution of rotation periods in NGC 2362 that covers the range 0.3 to 30 days. This obtained distribution of rotation periods is plotted in the upper left graph of Fig. 3.

The evolution of each initial rotation period of this reference distribution was then calculated using the angular momentum evolution model described in Sect. 2 with the  parameter listed in Tab.1.  Figure 3 shows the simulated evolution of this 5 Myrs old reference distribution at subsequent ages of 13, 25, 130, 550, and 1250 Myrs. It indicates that during the first 13 Myrs the broad initial distribution is shifted towards shorter rotation periods due to the effect of stellar contraction. At an age of about 25 Myrs, a double peak distribution appears that lasts till the age of about 550 Myrs. At later ages, the simulation produces a single peak distribution that sharpens and moves with time towards longer rotation periods.

\begin{figure*}[!ht]
	\begin{tabular}{c c c}
	\centering
	\includegraphics[width=0.24\linewidth, angle=270]{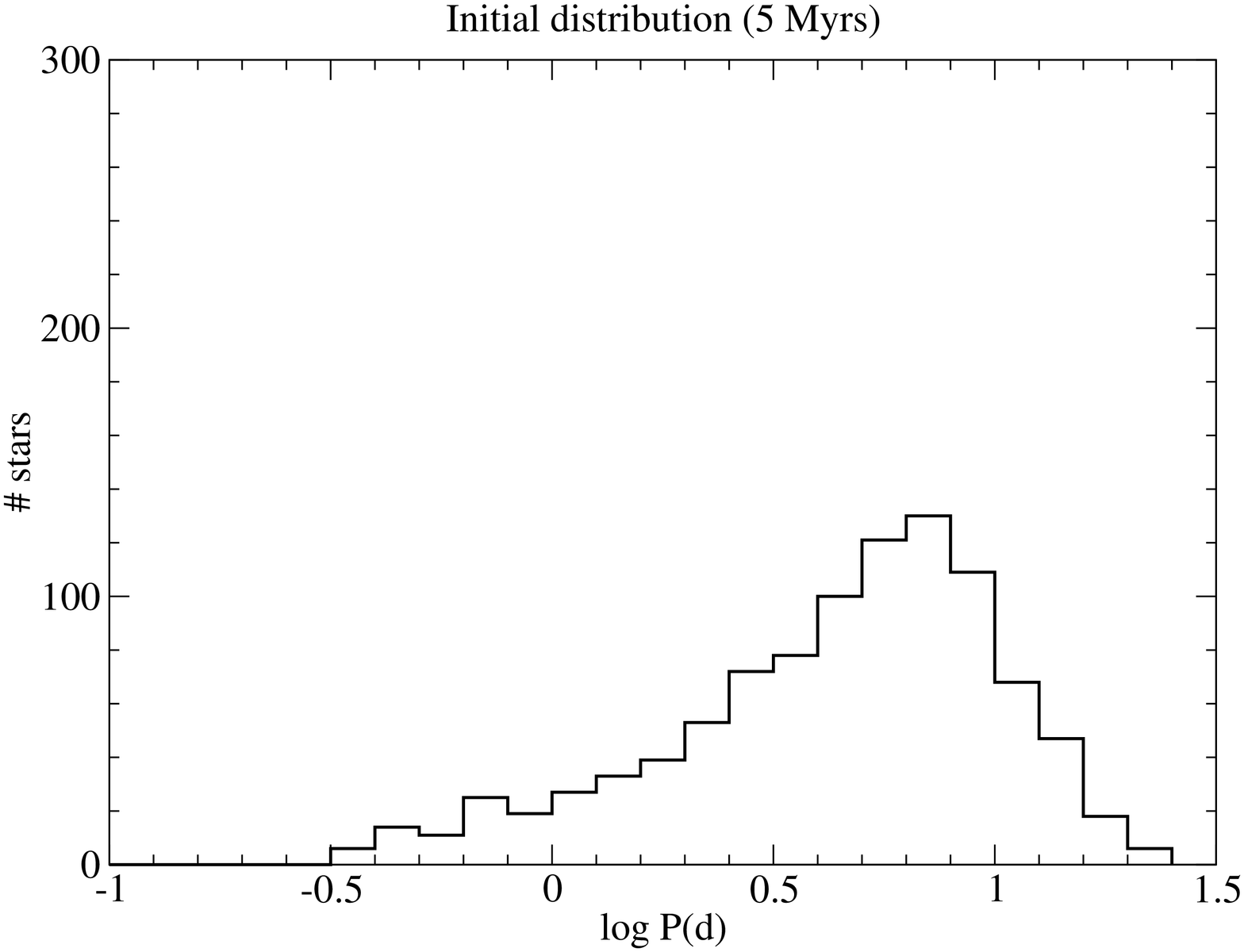} & \includegraphics[width=0.24\linewidth, angle=270]{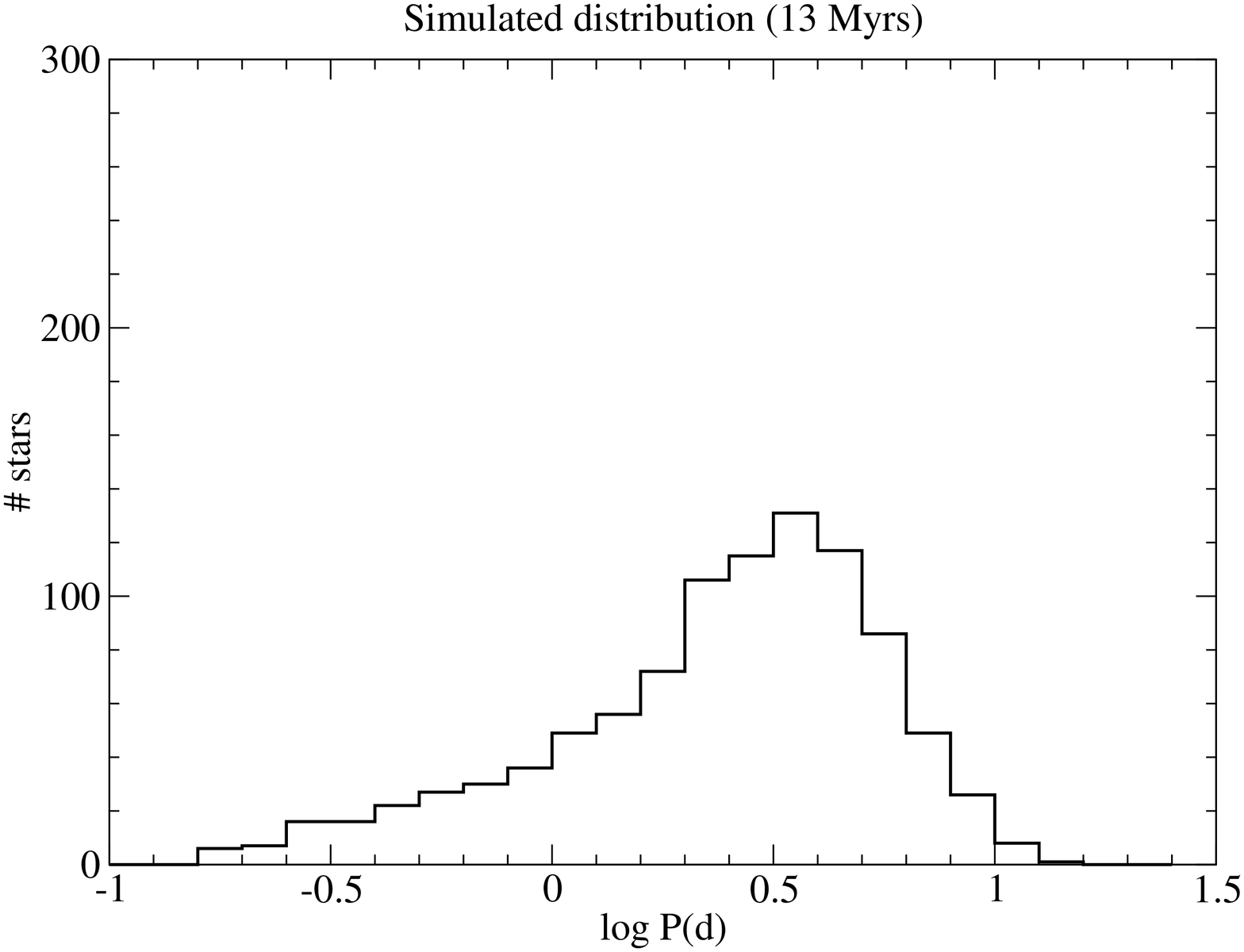} & \includegraphics[width=0.24\linewidth, angle=270]{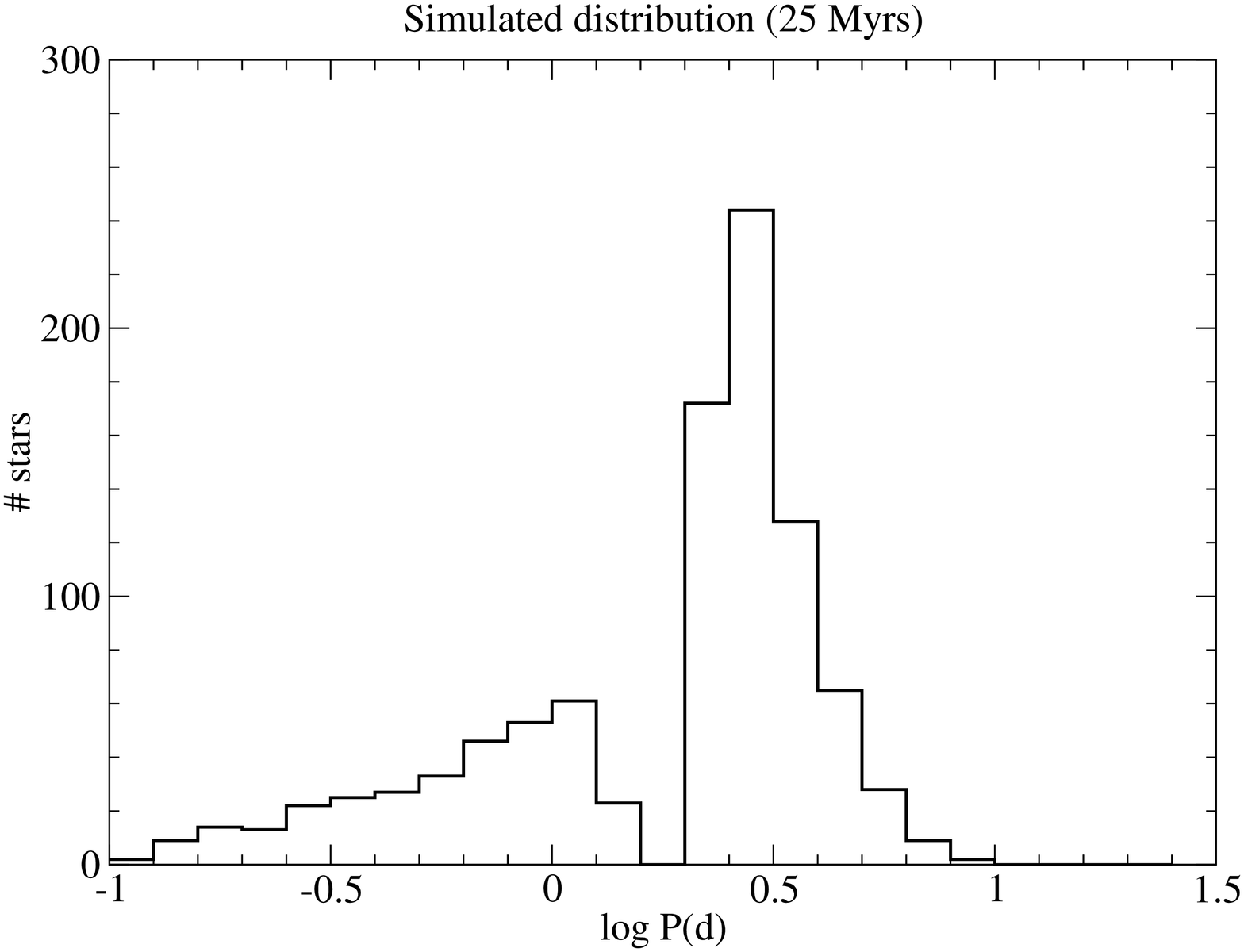} \\
	\includegraphics[width=0.24\linewidth, angle=270]{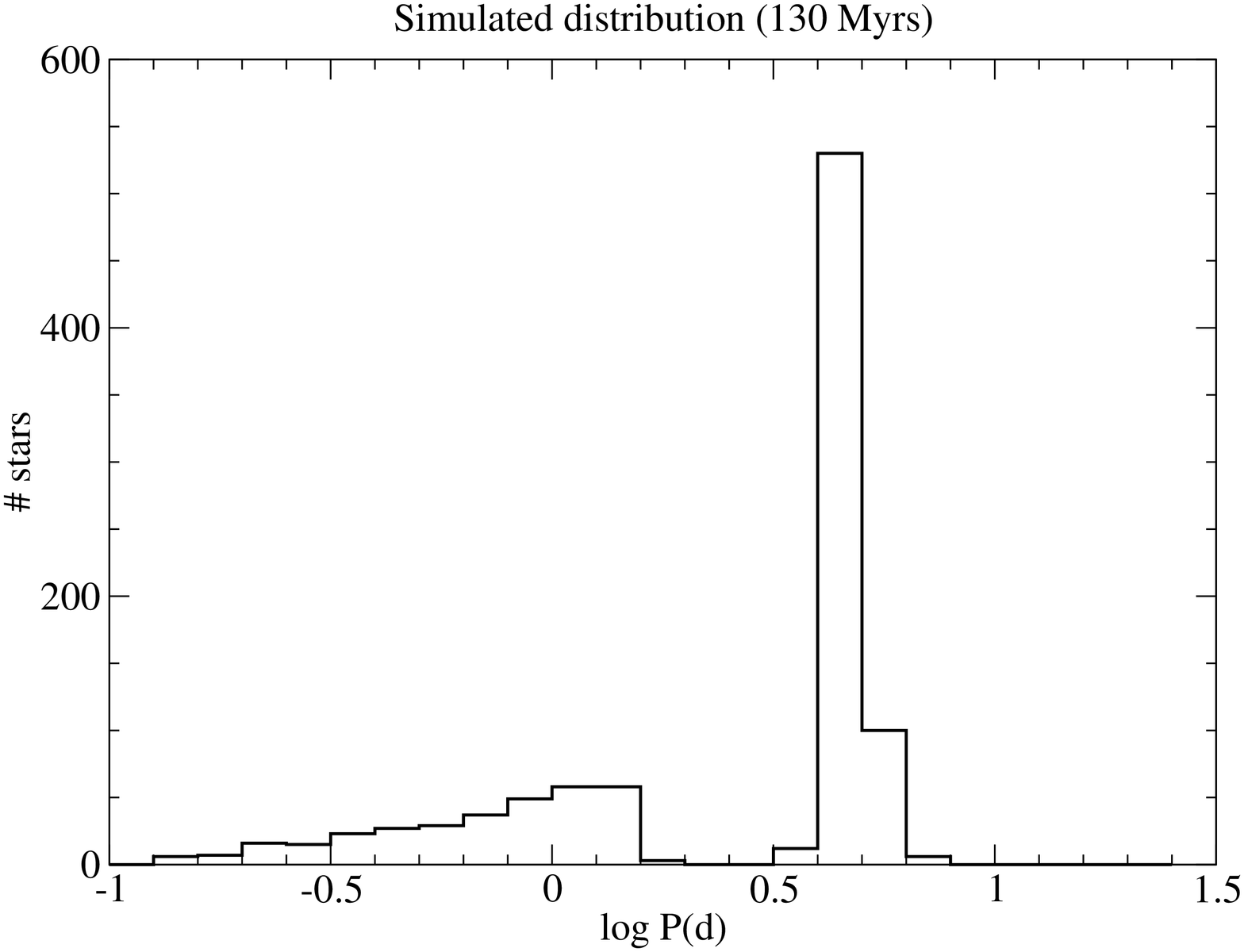} & \includegraphics[width=0.24\linewidth, angle=270]{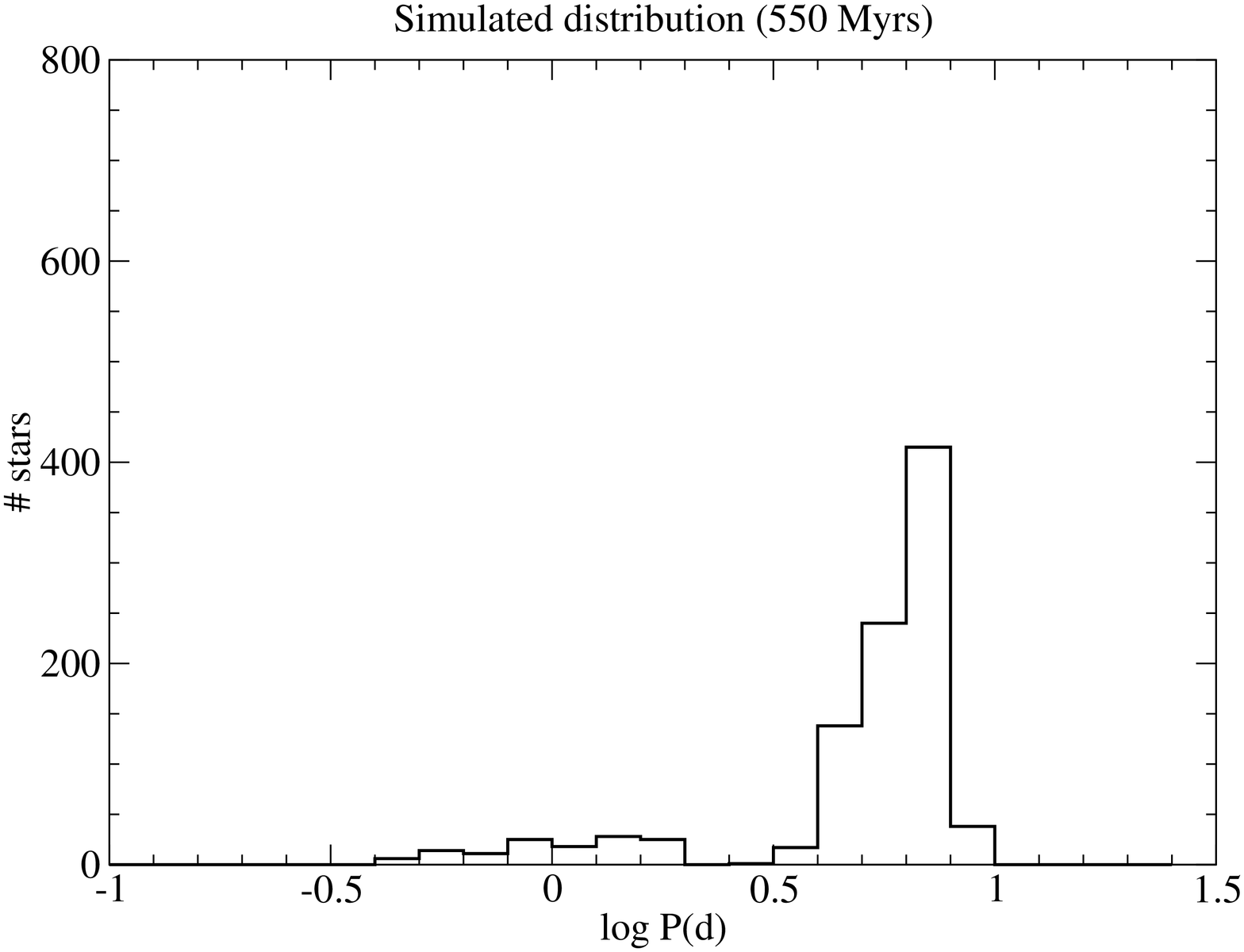} & \includegraphics[width=0.24\linewidth, angle=270]{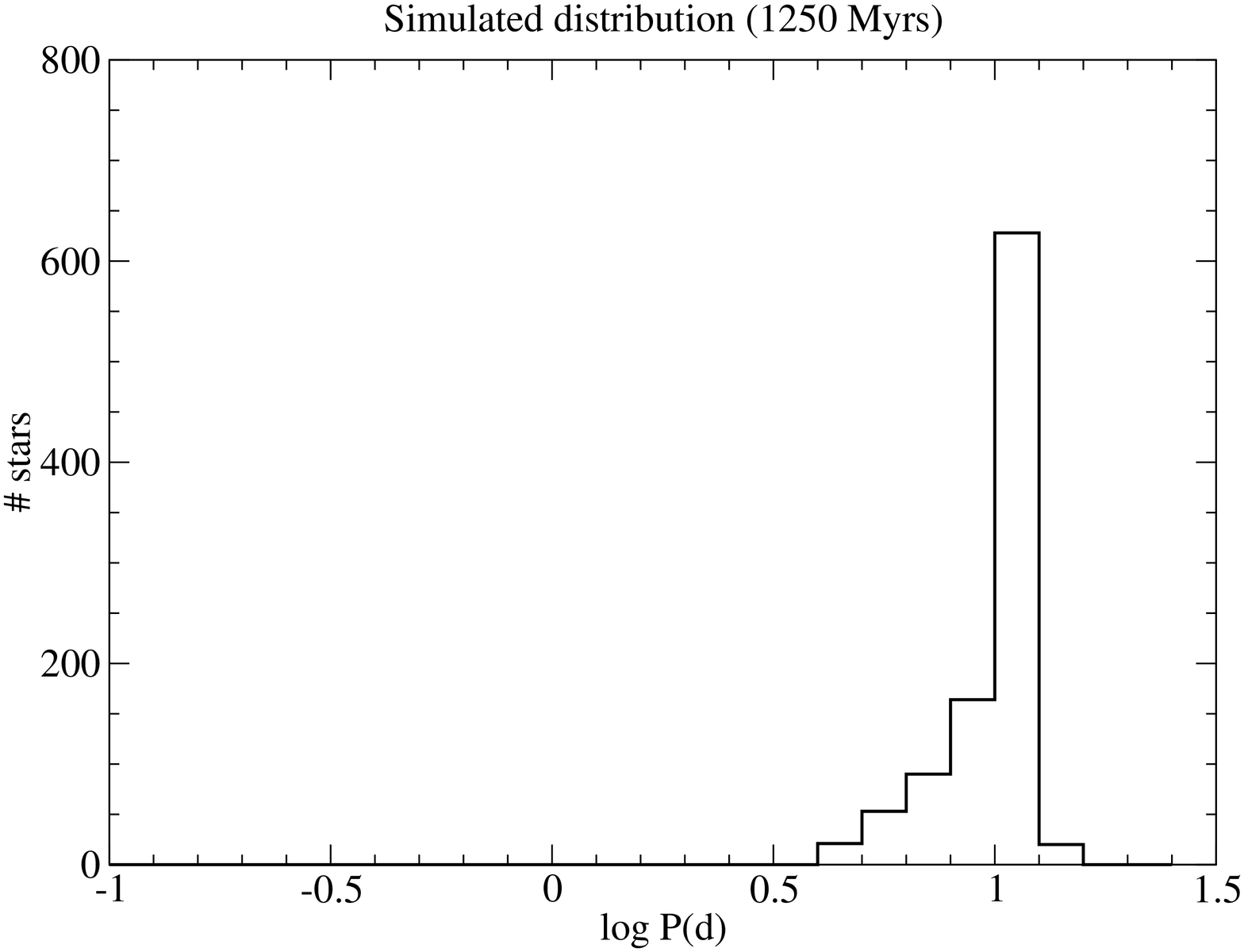} \\	
	\end{tabular}
	\caption{Simulated evolution of a normal rotation period distribution (upper left) similar to that of NGC 2362 (see text) at subsequent ages of 13 Myrs (upper middle), 25 Myrs (upper right), 130 Myrs (lower left) , 550 Myrs (lower middle), and 1.25 Gyrs (lower right).}
	\label{fig:fig_meas_rot_distrib}
\end{figure*}

\section{Comparison with observations}

\begin{figure*}
	\begin{tabular}{c c c}
	\centering
	\includegraphics[width=0.25\linewidth, angle=270]{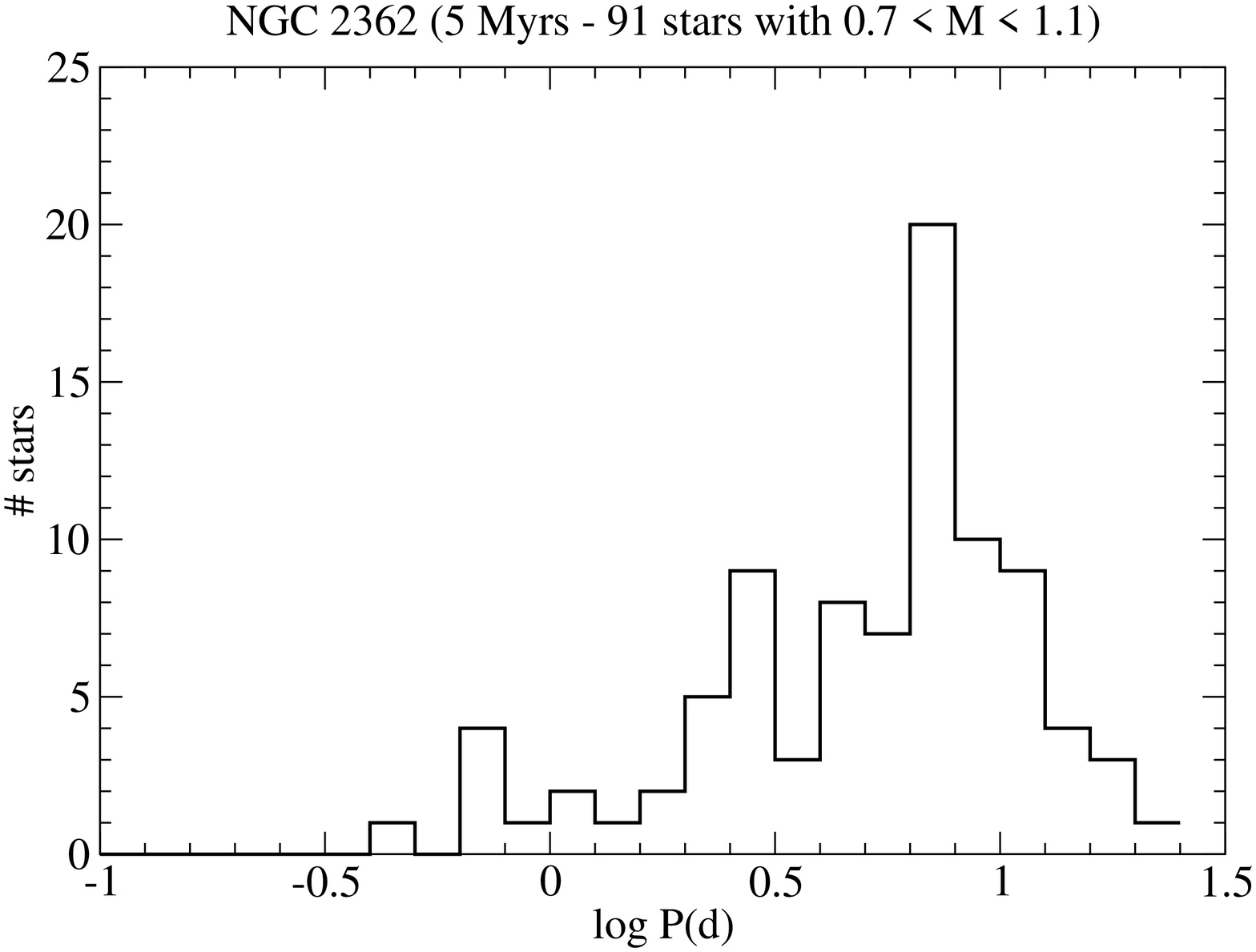} & \includegraphics[width=0.25\linewidth, angle=270]{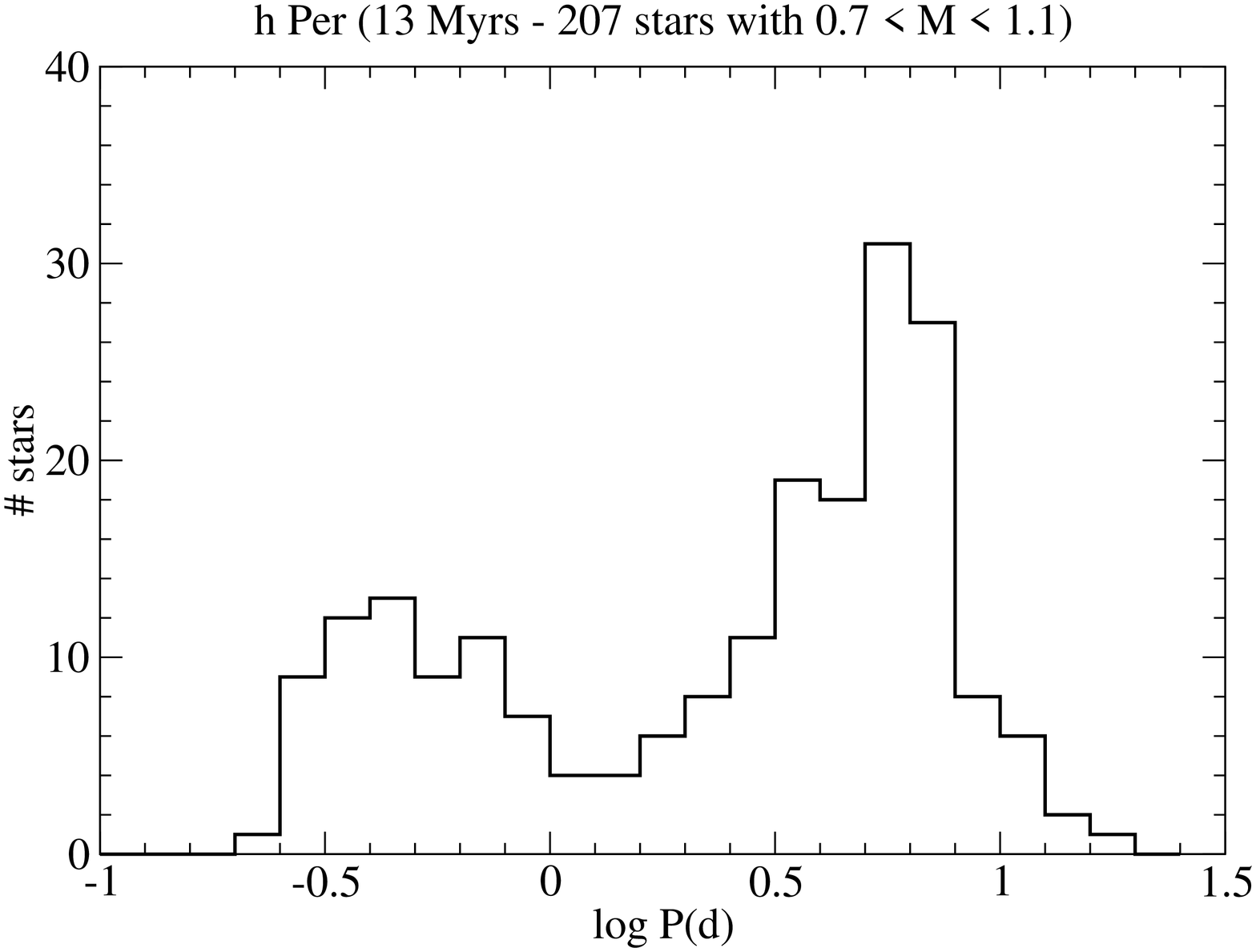} & \includegraphics[width=0.25\linewidth, angle=270]{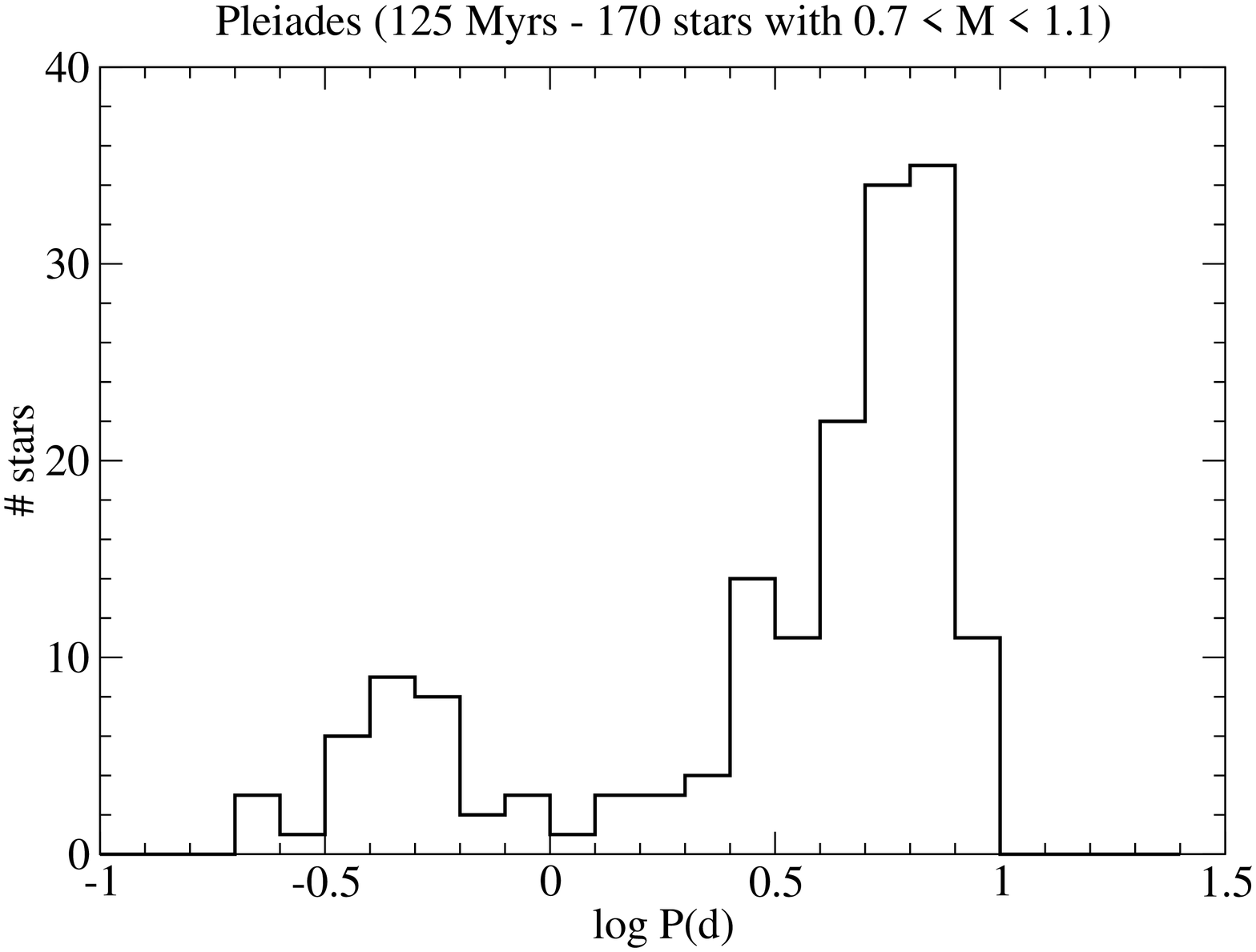} \\
	\includegraphics[width=0.25\linewidth, angle=270]{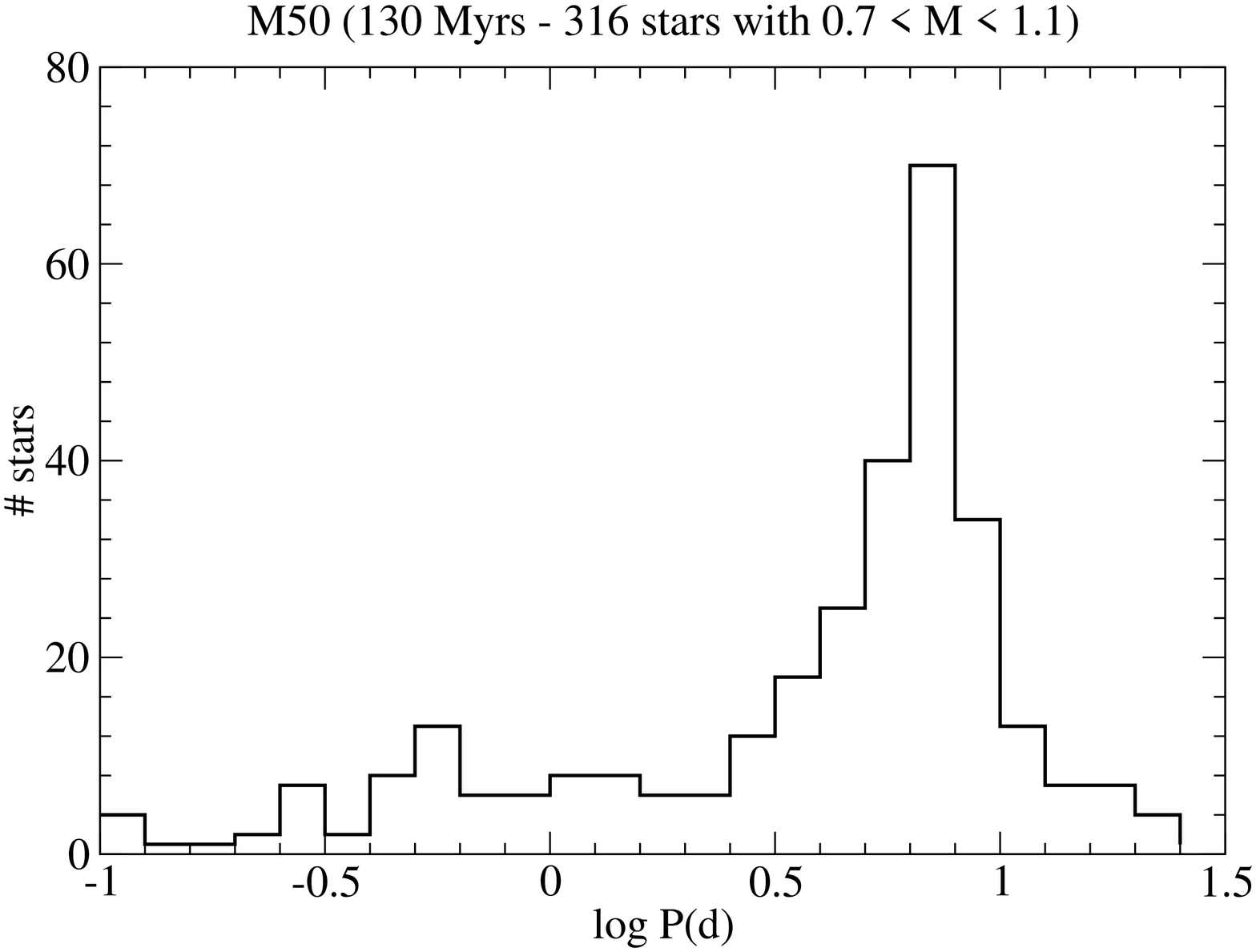} & \includegraphics[width=0.25\linewidth, angle=270]{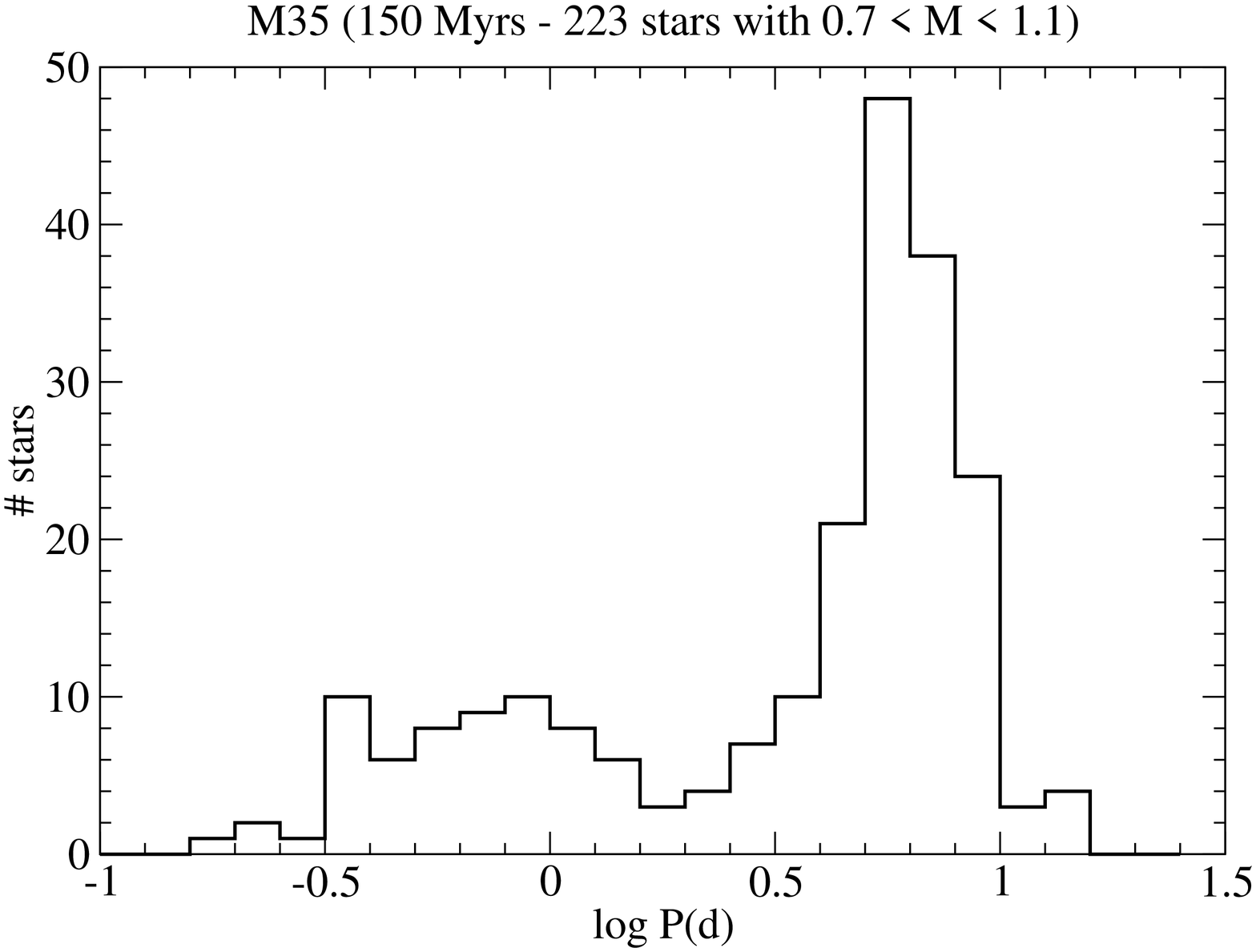} & \includegraphics[width=0.25\linewidth, angle=270]{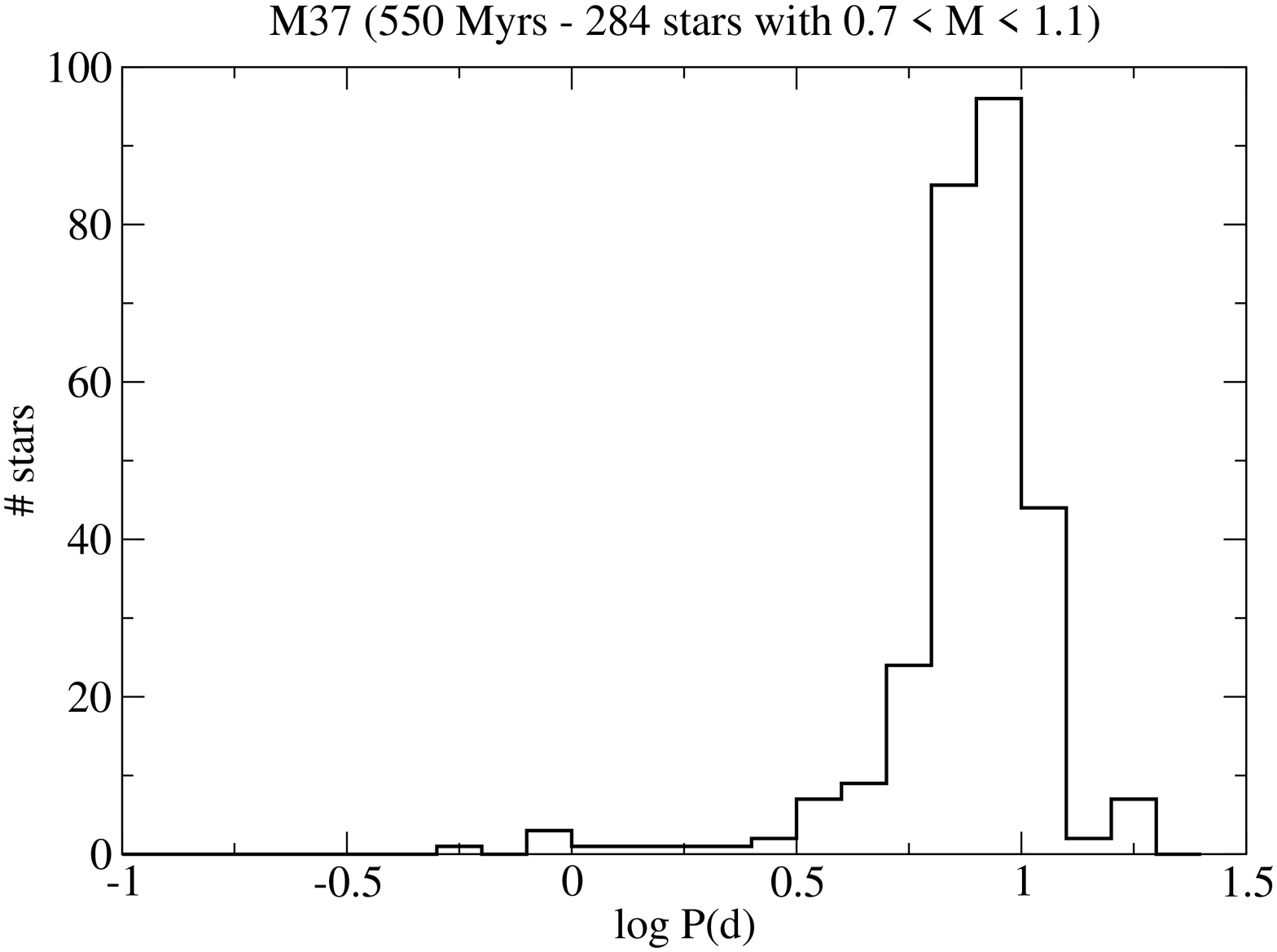} \\
	\end{tabular}
	\caption{Measured rotation period histogram of 0.7 - 1.1 M$_{\odot}$ stars in NGC 2362 (5 Myrs; Irwin et al. 2008), h Per (13 Myrs; Moraux et al. 2013), Pleiades (125 Myrs; Hartman et al. 2010), M50 (130 Myrs; Irwin et al. 2009), M35 (150 Myrs; Meibom et al. 2009), and M37 (550 Myrs; Hartman et al. 2009).}
	\label{fig:fig_meas_rot_distrib}
\end{figure*}

In order to compare the simulated evolution of the reference NGC 2362-like distribution with observations, I look for open clusters having a large enough number of Sun-like stars with known rotation periods. Open clusters with hundred such stars are estimated to be of a reasonable size for establishing rotation period distributions with large enough statistical significance. I initially defined Sun-like stars as stars within the 0.9 to 1.1 M$_{\odot}$ mass range. However, since I could find large enough samples in only two clusters, namely  h Per and M 35, I extended the mass bin to the 0.7 - 1.1 M$_{\odot}$ range. Three additional clusters were found having large enough samples of 0.7 - 1.1 M$_{\odot}$ stars with known rotation periods, namely the Pleiades, M50, and M37. The names of these clusters, their ages, the number of sample stars, and the references to the parent surveys are given in Table 2.

\begin{table}[!b]
	\centering
	\caption{Large samples of 0.7 - 1.1 M$_{\odot}$ stars with measured rotation periods used for comparisons with simulated rotation period distributions.}
	\label{tab:table_meas_rot_distrib}
	\begin{tabular}{c c c c}
	\noalign{\smallskip}\hline\hline\noalign{\smallskip}
	Cluster    &  Age  & $N_{\rm stars}$ & Reference \\
			& (Myrs) & & \\
	\noalign{\smallskip}\hline\noalign{\smallskip}
	NGC 2362 & 5 & 91 & Irwin et al. 2008 \\
	h Per & 13 & 207 & Moraux et al. 2013 \\
	Pleiades & 125 & 170 & Hartman et al. 2010 \\
	M50 & 130 & 316 & Irwin et al. 2009 \\
	M35 & 133 & 223 & Meibom et al. 2009 \\
	M37 & 550 & 284 & Hartman et al. 2009 \\	
	\noalign{\smallskip}\hline
	\end{tabular}
\end{table}

The study assumes that the rotation period distributions derived from the photometric surveys are representative of the cluster distribution and not significantly affected by observational biases.  Figure 4 shows the rotation period histograms of the 0.7 - 1.1 M$_{\odot}$ stars extracted from the selected surveys. A logarithmic scale is used on the x-axis such that the shapes of the histograms are the same whether the x-axis uses rotation periods or angular velocities bins.

The upper middle graph in Fig.4 shows the measured rotation period distribution of 0.7 - 1.1 M$_{\odot}$ stars in h Per (NGC 869) which age has been estimated to 13 Myrs (Mayne \& Naylor 2008). By the age of this cluster, the disk accretion process has ceased on all stars that freely evolve towards the ZAMS. The h Per sample shows a bi-modal distribution with a fast rotator group having rotation periods around 0.3-0.4 days, a slow rotator group with periods around 5-6 days, and few stars within the intermediate 1-2 days period range. The simulated rotation evolution of the NGC 2362-like reference sample does not show a bimodal distribution by the age 13 Myrs (see upper middle graph in Fig. 3) as observed in h Per but only 10 Myrs later (see upper right graph in Fig. 3).

The Pleiades, M50, and M35 open clusters have similar ages estimated to $\sim$125 Mys (e.g. Stauffer et al. 1998), $\sim$130 Myrs (e.g. Kalirai et al. 2003), and $\sim$133 Myrs (e.g. McNamara et al. 2011) respectively. They have been the subject of extensive photometric time-series survey by Hartman et al. (2010), Irwin et al. (2009), and Meibom et al. (2009), respectively. The three stellar samples show similar rotation period distributions (see upper right, lower left, and lower middle graphs in Fig.4). These include an important group of slow rotators with periods ranging between 2.5 and 10 days but concentrated around 5-8 days.  A smaller group of fast rotators is also visible between 0.25 and 1.5 days in the Pleiades and M35 histograms. In M50, the fast rotators group appears as a long tail in the rotation period distribution. The simulation reproduces reasonably well the bimodal distribution of rotation periods in the Pleiades, and M35 with 32 \% of the stars in the fast rotator group (see lower left graph in Fig. 3).

The M37 open cluster has an age of 550 Myrs (Hartman et al. 2008). The distribution of rotation periods in M37 is narrower than in younger clusters. The histogram exhibits a strong peak of slow rotators at period between 6 and 10 days that is well reproduced by the simulation (see lower middle graph in Fig. 3). This distribution is consistent with the view that stars in old clusters such as M37 and the Hyades are slow rotators while younger open clusters such as the Pleiades, M50 and M35 show a wide spread distribution of angular rotation. The bimodal distribution of stellar rotation observed in those clusters has almost fully disappeared in M37. This evolutionary trend is accounted for in the simulated evolution of the NGC 2362-like rotation period distribution (see Fig. 3). 

\section{Discussion}

I used a phenomenological model of angular momentum redistribution to simulate the evolution of a normal distribution of rotation velocities that emulates the distribution of rotation periods in the young NGC 2362 open cluster. The simulation results reproduces the main features of the rotation period distributions observed in older clusters.  They account for the  bimodal distribution of stellar rotation that is observed among Sun-like stars in the Pleiades, M50 and M35.  The model simulate the prospective appearance of a bimodal rotation period distribution at an early stage of NGC 2362 evolution, though not as early as observed in the h Per open cluster.  These results suggest that the stellar rotation periods distributions in open clusters form a continuous sequence of evolution. Clusters of similar ages have similar rotation periods distributions.  

One major event in the evolution of stellar rotation is needed by the model to reproduce the observed appearance and disappearance of the bimodal distribution.  This event is a brief episode of large angular momentum loss that would occur as the rotation rate of Sun-like stars decays through the 0.13 - 0.3 Rossby number interval (see Fig. 1). The enhanced braking torque induces a strong deceleration of stellar rotation that explains the rapid transition of Sun-like stars from the fast to the slow rotators sequence observed in young open clusters (Gondoin 2012, 2015). This catastrophic event occurs at ages included between a few tens and $\sim$ 600 Myrs depending on the initial rotation rate of the stars after circumstellar disk dispersion, thus accounting  for the bimodal distribution of stellar rotation observed in clusters within this range of ages.

The model assumes that this brief episode of enhanced angular momentum loss is due to a sudden rise of the mass loss rate at a critical rotation rate combined with a continuous and moderate increase of the Alfven radius with decreasing rotation rate. This scenario is suggested by measurements of astrospheric Ly$\alpha$ absorptions interpreted with the support of hydrodynamic models (Wood et al. 2005, 2014). These observations indicate that the mass loss rates of G-K dwarfs increases with the X-ray surface flux up to a wind dividing line beyond which the mass loss rate of active young stars would be significantly lower.  

Although the model assumes that the episode of enhanced rotational braking is driven by a sudden increase of the mass loss rate, alternative scenarios producing a similar evolution of the angular momentum loss are a priori equally plausible. This includes e.g the sudden change in the coupling between the wind and the large scale magnetic fields of the stars proposed in the metastable dynamo model described by Brown (2014). This scenario, however, does not seems to be consistent with spectro-polarimetric observations of solar-type stars (Donati \& Landstreet 2009). Folsom et al. (2016) found no strong difference in their sample of young Sun-like stars between the magnetic geometry of fast rotators (P $<$ 2 days) and that of moderate rotators, which would essentially correspond to the transition between the fast and slow rotation sequence of Barnes (2003). Their observations rather indicate a continuous evolution from strong, toroidal and non-axisymmetric magnetic fields on stars with small Rossby number to a more poloidal and axisymmetric configuration of large scale magnetic fields on moderate rotators.

An episode of enhanced angular momentum loss due to a more efficient magnetic braking by stellar winds should increase the differential rotation between the core and the envelope of the star. Since the magnetic field lines that sling charged particles from the wind into space are rooted in the photosphere, a strong wind torque is expected to decelerate the envelope rotation while the conservation of angular momentum should keep the radiative core in rapid rotation.  A large shear should develop at the base of the convection zone and trigger various instabilities. These instabilities are expected to drive mass motions or gravity waves that redistribute angular momentum and mix the stellar material enhancing light-element depletion  (e.g. Chaboyer et al. 1995; Charbonnel \& Talon 2005; Talon 2008).  Such an effect was observed in the Pleaides and M34 by Gondoin (2014) who noted that K stars on the fast rotator sequence have significantly higher lithium abundances that stars with same masses and ages located on the slow rotator sequence. This observed depletion of Li among stars that evolve from the fast to the slow rotator sequence supports the occurrence of a brief episode of enhanced magnetic braking by stellar wind in the early evolution of Sun-like stars. 

The possible causes of such a catastrophic event remain to be determined.  Remarkably, the rapid evolution of the stellar rotation in 0.13-0.3 Rossby number range and age domain seems to be correlated with a change in the evolution of the stellar X-ray emission level.  A steep transition in X-ray to bolometric luminosity ratio has been observed in the M34 open cluster (Gondoin 2012) between stars on the fast rotator sequence that emit close to the 10$^{-3}$ saturation level, and stars on the slow rotator sequence, whose $L_{\rm X}/L_{\rm bol}$ ratio is significantly lower.  Correlated transitions between the saturated and non-saturated X-ray emission regimes and between the fast and slow rotator sequences are also observed in the Pleiades (Gondoin 2015).  Based on these observations, Gondoin (2013) argued that the transition from the saturated to the non-saturated regime of X-ray emission among main-sequence stars may be the result of a dynamo regime transition.

\section{Conclusion}

The appearance of a bimodal distribution of rotation periods in young open clusters and the correlation of the rotation sequences with X-ray emission and Li abundance point towards a scenario where Sun-like stars with a rapid enough rotation after circumstellar disk dispersion experience a short episode of large rotational braking in their early evolution. This catastrophic event is driven by a sudden increase of the mass loss rate due to stellar winds at Rossby number included between 0.13 and 0.3. The resulting increase of the braking torque induces a large rotational shear at the bottom of the convective zone. It occurs on stars with ages included between 20-30 Myrs and $\sim$ 600 Myrs depending on their initial rotation rate after dispersion of their circumstellar disk,  thus accounting  for the bimodal distribution of stellar rotation observed in clusters with those ages. 

\section*{Acknowledgments}
{I am grateful to the organizing committees of the "Cool Star 19" workshop for allowing me to present this work.}


\begin{thebibliography}{}

\bibitem[\protect\citeauthoryear{Barnes}{2003}]{Barnes2003a} Barnes, S. A. 2003, \apj, 586, 464
\bibitem[\protect\citeauthoryear{Brown}{2014}]{Brown2014} Brown, T. M. 2014, \apj, 796, 91
\bibitem[\protect\citeauthoryear{Chaboyer et al.}{1995}]{Chaboyer1995} Chaboyer, B., Demarque, P., \& Pinsonneault, M. H. 1995, \apj, 441, 876
\bibitem[\protect\citeauthoryear{Charbonnel et al.}{2005}]{Charbonnel2005} Charbonnel, C. \& Talon, S. 2005, Science, Vol. 309, Iss. 5744, p. 2189
\bibitem[\protect\citeauthoryear{Donati & Landstreet}{2009}]{Donati2009} Donati, J.-F., \& Landstreet, J. D. 2009, \araa, 47, 333
\bibitem[\protect\citeauthoryear{Folsom et al.}{2016}]{Folsom2016} Folsom, C. P., Petit, P., Bouvier, J. et al. 2016, \mnras, 457, 580
\bibitem[\protect\citeauthoryear{Gondoin}{2012}]{Gondoin2012b} Gondoin, P. 2012, \aap, 546, A117
\bibitem[\protect\citeauthoryear{Gondoin}{2013}]{Gondoin2013} Gondoin, P. 2013, \aap, 556, A14
\bibitem[\protect\citeauthoryear{Gondoin}{2014}]{Gondoin2014} Gondoin, P. 2014, \aap, 566, A72
\bibitem[\protect\citeauthoryear{Gondoin}{2015}]{Gondoin2015} Gondoin, P. 2015, Proceedings of the 2nd Solarnet meeting: solar and stellar magnetic activity, http://www.astropa.inaf.it /Solarnet2015/Proceedings/ Proceedings.html
\bibitem[\protect\citeauthoryear{Hartman et al.}{2008}]{Hartman2008} Hartman, J. D., Gaudi, B. S., Holman M. J. et al. 2008, \apj, 675, 1254
\bibitem[\protect\citeauthoryear{Hartman et al.}{2009}]{Hartman2009} Hartman, J. D., Gaudi, B. S., Pinsonneault, M. H. et al. 2009, \apj, 691, 342
\bibitem[\protect\citeauthoryear{Hartman et al.}{2010}]{Hartman2010} Hartman, J. D., Bakos, G. A., Kovacs, G., \& Noyes, R. W. 2010, \mnras, 408, 475
\bibitem[\protect\citeauthoryear{Irwin et al.}{2008}]{Irwin2008} Irwin, J., Hodgkin, S., Aigrain, S. et al. 2008, \mnras, 384, 675
\bibitem[\protect\citeauthoryear{Irwin et al.}{2009}]{Irwin2009} Irwin, J., Aigrain, S., Bouvier, J. et al. 2009, \mnras, 392, 1456
\bibitem[\protect\citeauthoryear{Judge}{2003}]{Judge2003} Judge, P.G., Solomon, S. C., \& Ayres, T. R. 2003, \apj, 593, 534
\bibitem[\protect\citeauthoryear{Kalirai}{2003}]{Kalirai2003} Kalirai, J. S., Fahlman G. G., Richer, H. B., \& Ventura, P. 2003, \aj, 126, 1402
\bibitem[\protect\citeauthoryear{McNamara et al.}{2011}]{McNamara2011} McNamara, B. J., Harrison, T. E., McArthur, B. E., \& Benedict, G. F. 2011, \aj 142, 53
\bibitem[\protect\citeauthoryear{Mayne et al.}{2007}]{Mayne2007} Mayne N.J., Naylor, T., Littlefair S. et al. 2007, \mnras, 375, 1220
\bibitem[\protect\citeauthoryear{Mayne & Naylor}{2008}]{Mayne2008} Mayne N.J. \& Naylor, T. 2008, \mnras, 386, 261
\bibitem[\protect\citeauthoryear{Meibom et al.}{2009}]{Meibom2009} Meibom, S., Mathieu, R. D., \& Stassun, K. G. 2009, \apj, 695, 679
\bibitem[\protect\citeauthoryear{Meibom et al.}{2011}]{Meibom2011} Meibom, S., Matthieu, R. D., Stassun, K. G. et al. 2011, \apj, 733, 115
\bibitem[\protect\citeauthoryear{Moraux et al.}{2013}]{Moraux2013} Moraux, E., Artemenko, S., Bouvier, J. et al. 2013, \aap, 560, A13
\bibitem[\protect\citeauthoryear{Oglethorpe & Garaud}{2013}]{Oglethorpe2013} Oglethorpe, R. L. F., \& Garaud, P. 2013, \apj, 778, 166
\bibitem[\protect\citeauthoryear{Pizzolato et al.}{2003}]{Pizzolato2003} Pizzolato, N., Maggio, A., Micela, G., Sciortino, S., \& Ventura, P. 2003, \aap, 397, 147
\bibitem[\protect\citeauthoryear{Siess et al.}{2000}]{Siess2000} Siess, L., Dufour, E., \& Forestini, M.  2000, \aap, 358, 593
\bibitem[\protect\citeauthoryear{Skumanich et al.}{1972}]{Skumanich1972} Skumanich, A. 1972, \apj, 171, 565
\bibitem[\protect\citeauthoryear{Spada et al.}{2011}]{Spada2011} Spada, F., Lanzafame, A.C., Lanza, A.F. et al. 2011, \mnras, 416, 447
\bibitem[\protect\citeauthoryear{Stauffer et al.}{1998}]{Stauffer1998} Stauffer, J. R., Schild, R., Barrado y Navascues, D. et al. 1998, \apj, 504, 805 
\bibitem[\protect\citeauthoryear{Talon}{2008}]{Talon2008} Talon, S. 2008, Memorie della Societa Astronomica Italiana, 79, 569 
\bibitem[\protect\citeauthoryear{Weber \& Davis}{1967}]{Weber1967} Weber, E.J, \& Davis, L., Jr. 1967, \apj, 148, 217
\bibitem[\protect\citeauthoryear{Wood et al.}{2005}]{Wood2005} Wood, B. E., M\"uller, H.-R., Zank, G. P. et al. 2005, \apj, 628, L143
\bibitem[\protect\citeauthoryear{Wood et al.}{2014}]{Wood2014} Wood, B. E., M\"uller, H.-R., Redfield, S., \& Edelaman, E. 2014, \apjl, 781, L33
\bibitem[\protect\citeauthoryear{Wright et al.}{2011}]{Wright2011} Wright, N. J., Drake, J. J., Mamajek, E. E., \& Henry, G. W. 2011, \apj, 743, 48 

\end{thebibliography}

\newpage

\end{document}